\documentclass[aps,prb,twocolumn,showpacs,groupedaddress]{revtex4-2}
\usepackage{graphicx}
\usepackage{epstopdf} 
\begin{document}

\title{Theory of light reflection and transmission by a plasmonic nanocomposite slab: Emergence of broadband perfect absorption}

\author{V.G.~Bordo}
\email{bordo@mci.sdu.dk}

\affiliation{SDU Electrical Engineering, University of Southern Denmark, Alsion 2, DK-6400 S{\o}nderborg, Denmark}


\date{\today}

\begin{abstract}
A theory of light reflection and transmission by an optically thin nanocomposite slab which contains randomly distributed metal nanoparticles (NPs) is developed. The underlying model takes into account the reflection of light scattered by NPs from the slab boundaries, enhanced decay of localized surface plasmons in dense NP arrays and light scattering at the slab surface -- the factors which are beyond the scope of the Maxwell Garnett approximation. It is demonstrated that the first two effects lead to broadband perfect absorption observed in such nanocomposites, whereas the last one is responsible for its omnidirectional character and polarization insensitivity. These findings open up new possibilities for engineering broadband perfect absorption in plasmonic nanocomposites.
\end{abstract}


\maketitle

\section{Introduction}
Efficient broadband optical absorption is required in diverse applications, in particular, in solar cells, thermal sources and bolometers. On the other hand, the incident light being almost completely absorbed by a structure is faintly reflected that is an advantage for anti-reflective coatings which are in high demand, for example, for contrast improvement in optical instruments due to the elimination of stray light. A traditional approach to the latter problem is exploiting of either refractive index matching coatings or multi-layer films which provide destructive interference in reflection \cite{Born70}. In both cases the structure of the coating is dictated by the wavelength of the incident light and is therefore wavelength-specific.\\
Plasmonic metamaterials, which are artificial structures containing nano-sized metallic inclusions, suggest a novel solution in this field. Metal nanoparticles (NPs) support collective electron oscillations known as localized surface plasmons (LSPs) which have broad resonances in the optical spectral range. NPs can be arranged at a metal film, either regularly or randomly, being separated from it by a spacer layer (film-coupled nanoantennas) \cite{Soukoulis09,Qiu10,Giessen10,Huang11,Qiu11,Savoy11,Smith12,Smith15,Smith16}. In such a case the gaps between NPs and the film support nanocavity plasmonic resonances which participate in the optical response of the structure as well. These resonances are dictated by the geometry and dimensions of the metamaterial unit cell and are responsible for narrow-band perfect absorption. A different metamaterial design which ensures a broadband efficient absorption contains a nanostructured metal film instead of patch nanoantennas \cite{Grigorenko08,Atwater11}.\\
An alternative approach to plasmonic perfect absorbers is based on nanocomposite ultra-thin films in which metal NPs are distributed randomly throughout the host dielectric in the course of the fabrication process \cite{Grigorenko10,Elbahri11,Elbahri12,Tamulevicius18}. Such structures provide absorption at the level of above 90\% in a wide spectral range for a broad interval of angles of incidence. Besides that, their absorption spectra weakly depend on the angle of incidence and polarization of the illuminating light. In this paper, we focus ourselves on this kind of plasmonic perfect absorbers.\\
To explain their results, the authors of Ref. \cite{Grigorenko10} assumed that the optical response of the Ag-Al$_2$O$_3$ nanocomposite film is the same as that of an Ag film of thickness equal to the Ag NP diameter and introduced an effective complex refractive index deduced from ellipsometric measurements with a silver film. This approach allowed one to reproduce large values of the absorption coefficient, however did not display its broadband character. The broadband perfect absorption observed in the nanocomposite structures containing Au and Cu NPs \cite{Elbahri11,Elbahri12} was attributed to both the coupling between the LSPs in metal NPs and propagating surface plasmon polaritons in the underlying metal film and to the resonances in the spacer layer between them, although no supporting calculations were provided. These papers noted, however, that reflections from the nanocomposite boundaries probably play a role in perfect absorption.\\ 
In order to consistently explain these observations and be able to engineer plasmonic perfect absorbers one needs a first-principles approach. Recently, we have developed a theory which rigorously treats the local field in a nanocomposite film of sub-wavelength thickness \cite{Bordo18}. We have shown that the account of the field scattered by NPs in the nanocomposite slab and reflected by its boundaries leads to anisotropy and nonlocality of the nanocomposite optical response. It was speculated that this effect may play a role in perfect absorption and that the light scattering at surface roughness can contribute to it.\\
In this paper, basing on a proper account of the local field in a plasmonic nanocomposite slab, we develop the theory of light reflection and transmission by such a slab. We derive analytical expressions for the reflection and transmission coefficients which should be used for such a structure instead of the Fresnel equations. To match the experimental conditions we develop a model which takes into account surface roughness and investigate its impact on the reflection spectrum. A special attention is paid to the enhancement of the relaxation processes which disrupt the LSP oscillations in dense NP arrays - the mechanisms which, to the best of our knowledge, have not been discussed in the literature on the effective medium theory before. \\
The paper is organized as follows. In Sec. \ref{sec:local} we briefly review the local field approach which was first developed in Ref. \cite{Bordo18} and extends the Maxwell Garnett approximation to optically thin nanocomposites. In Sec. \ref{sec:refl_trans} we derive the reflection and transmission coefficients for a slab for both TM and TE polarizations which take into account light scattering at surface roughness. Some numerical results which illustrate the theory as well as their comparison with the Maxwell Garnett and Bruggeman approximations are given in Sec. \ref{sec:numer}. Section \ref{sec:concl} summarizes the main results of the paper.
\section{Local field in a slab}\label{sec:local}
\begin{figure}
\includegraphics[width=\linewidth]{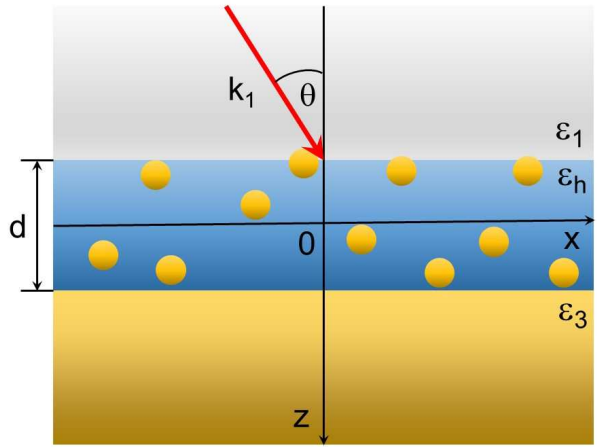}
\caption{\label{fig:geometry} Geometry of the problem.}
\end{figure} 
We consider an infinite nanocomposite slab of thickness $d$ (medium 2) which is enclosed between semi-infinite media 1 and 3 with the dielectric functions $\epsilon_1$ and $\epsilon_3$, respectively (Fig. \ref{fig:geometry}). We direct the $z$ coordinate axis from medium 1 to medium 3 along the normal to the slab boundaries and place its origin at the midpoint between them. We assume that the slab contains spherical metal nanoparticles of radius $R$ randomly distributed with the number density $N$, the dielectric function of the host material is $\epsilon_h$ and the dielectric function of NPs is described by the Drude model
\begin{equation}\label{eq:drude}
\epsilon(\omega)=\epsilon_{\infty}-\frac{\omega_p^2}{\omega(\omega+i\Gamma_{NP})},
\end{equation}
where $\omega=2\pi c/\lambda$ is the frequency of the incident light with $c$ being the speed of light in vacuum and $\lambda$ being the wavelength, $\epsilon_{\infty}$ is the offset originating from the interband transitions, $\omega_p$ is the metal plasma frequency and $\Gamma_{NP}$ is the relaxation constant. \\
We follow Maxwell Garnett's assumption \cite{MG} that the metal spheres having the radius much smaller than the wavelength of the incident light ($R\ll\lambda$) can be replaced by point dipoles with the dipole moments 
\begin{equation}\label{eq:polarization}
{\bf p}({\bf r},\omega)=\alpha(\omega){\bf E}^{\prime}({\bf r},\omega),
\end{equation}
where
\begin{equation}
\alpha(\omega)=\epsilon_hR^3\frac{\epsilon(\omega)-\epsilon_h}{\epsilon(\omega)+2\epsilon_h}
\end{equation}
is the sphere polarizability in the quasistatic approximation and ${\bf E}^{\prime}({\bf r},\omega)$ is the microscopic (local) field acting on the sphere. Let us note that the zero of the denominator in $\alpha(\omega)$ determines the frequency of the LSP at which the optical response of NPs is maximum. The dipole moment of NPs in a unit volume of the nanocomposite, or their polarization, equals ${\bf P}({\bf r},\omega)=N{\bf p}({\bf r},\omega)$. \\
The local field in turn can be written in the form of an integral equation \cite{Born24,Bordo18}
\begin{equation}\label{eq:local}
{\bf E}^{\prime}({\bf r},\omega)={\bf E}({\bf r},\omega)+\int_{V^{\prime}} \bar{\bf F}({\bf r},{\bf r}^{\prime};\omega){\bf P}({\bf r}^{\prime},\omega)d{\bf r}^{\prime},
\end{equation}
which can be understood as a general solution of the microscopic Maxwell's equations represented as a sum of the general solution of the homogeneous Maxwell's equations, ${\bf E}({\bf r},\omega)$, and the partial solution of the inhomogeneous equations with the source given by the polarization ${\bf P}({\bf r},\omega)$. Here $\bar{\bf F}({\bf r},{\bf r}^{\prime};\omega)$ is the so-called field susceptibility tensor \cite{Sipe84,Nha96} which relates the electric field at the point ${\bf r}$ generated by a classical dipole, oscillating at frequency $\omega$, with the dipole moment itself, located at ${\bf r}^{\prime}$. This quantity can be expressed in terms of the Green's function of the vector wave equation with appropriate boundary conditions \cite{Bordo16}. The symbol $V^{\prime}$ in Eq. (\ref{eq:local}) denotes the slab volume after removal of a small volume around the NP under consideration that excludes the singularity associated with the self-action of NPs. It is a common procedure when one considers integral equations for the local field \cite{Born24}. When writing Eq. (\ref{eq:local}) one formally treats the quantity ${\bf P}({\bf r},\omega)$ as a continuous function of ${\bf r}$ that is a good approximation when the average distance between NPs is much less than $\lambda$. \\
The quantity $\bar{\bf F}({\bf r},{\bf r}^{\prime};\omega)$ can be split into two contributions as
\begin{equation}\label{eq:decompose}
\bar{\bf F}({\bf r},{\bf r}^{\prime};\omega)=\bar{\bf F}^0({\bf r},{\bf r}^{\prime};\omega)+\bar{\bf F}^R({\bf r},{\bf r}^{\prime};\omega),
\end{equation}
where $\bar{\bf F}^0({\bf r},{\bf r}^{\prime};\omega)$ originates from the dipole field of NPs and leads to the Lorentz local field \cite{Born70}. The other term, $\bar{\bf F}^R({\bf r},{\bf r}^{\prime};\omega)$, results from the NPs dipole field reflected from the boundaries and is found in Ref. \cite{Nha96}. It was first considered in the context of the local field theory in Ref. \cite{Bordo18}. The reflected contribution, $\bar{\bf F}^R({\bf r},{\bf r}^{\prime};\omega)$, decreases with the distance from the boundary and becomes insignificant at distances much larger than the wavelength. However, in a subwavelength slab its influence is substantial.\\
The field ${\bf E}({\bf r},\omega)$ in Eq. (\ref{eq:local}) corresponds to a solution for the slab without sources, i.e. with the NPs being removed. It can be therefore interpreted as the "incident" field in the slab which is not yet disturbed by NPs. Such a consideration is typical for a scattering problem. In the bulk medium, where the reflected contribution can be neglected, this field equals the mean field due to an isotropic distribution of the NP dipoles. However, in the near field of the boundaries or inside a subwavelength slab this equivalence is violated.\\
Introducing the spatial Fourier transforms
\begin{equation}\label{eq:Efourier}
{\bf E}({\bf r},\omega)=\frac{1}{(2\pi)^2}\int {\bf E}({\bf k}_{\parallel},\omega; z)e^{i{\bf k}_{\parallel}\cdot {\bf r}_{\parallel}}d{\bf k}_{\parallel},
\end{equation}
\begin{equation}\label{eq:Pfourier}
{\bf P}({\bf r},\omega)=\frac{1}{(2\pi)^2}\int {\bf P}({\bf k}_{\parallel},\omega; z)e^{i{\bf k}_{\parallel}\cdot {\bf r}_{\parallel}}d{\bf k}_{\parallel}
\end{equation}
and
\begin{equation}\label{eq:Ffourier}
\bar{\bf F}^R({\bf r},{\bf r}^{\prime};\omega)=\frac{1}{(2\pi)^2}\int \bar{\bf F}^R({\bf k}_{\parallel},\omega;z,z^{\prime})e^{i{\bf k}_{\parallel}\cdot ({\bf r}_{\parallel}-{\bf r}^{\prime}_{\parallel})}d{\bf k}_{\parallel}
\end{equation}
with ${\bf r}_{\parallel}$ and ${\bf k}_{\parallel}$ being the radius vector and wave vector components parallel to the slab boundaries, respectively, one obtains from Eqs. (\ref{eq:polarization}), (\ref{eq:local}) and (\ref{eq:decompose})
\begin{eqnarray}\label{eq:integral_fourier}
{\bf P}( k_{\parallel},\omega; z)
-\chi(\omega)\int_{-d/2}^{d/2} \bar{\bf F}^R( k_{\parallel},\omega ; z,z^{\prime}){\bf P}( k_{\parallel},\omega ; z^{\prime})dz^{\prime}\nonumber\\
=\chi(\omega){\bf E}( k_{\parallel},\omega ; z).\nonumber\\
\end{eqnarray}
Here
\begin{equation}\label{eq:chi}
\chi(\omega)=\frac{N\alpha(\omega)}{1-(4\pi/3\epsilon_h)N\alpha(\omega)}
\end{equation}
is the linear susceptibility of NPs \cite{Born70} and we have taken into account the Lorentz relation
\begin{equation}\label{eq:F0}
\int_{V^{\prime}} \bar{\bf F}^0({\bf r},{\bf r}^{\prime};\omega){\bf P}({\bf r}^{\prime},\omega)d{\bf r}^{\prime}=\frac{4\pi}{3\epsilon_h}{\bf P}({\bf r},\omega).
\end{equation}
In what follows we assume that the slab is optically thin so that $d/\lambda\ll 1$. In this case one can approximate the kernel in Eq. (\ref{eq:integral_fourier}) as 
\begin{equation}\label{eq:F0R}
\bar{\bf F}^R(k_{\parallel},\omega;z,z^{\prime})\approx \bar{\bf F}^R(k_{\parallel},\omega;0,0)\equiv \bar{\bf F}_0^R(k_{\parallel},\omega).
\end{equation}
As a result, Eq. (\ref{eq:integral_fourier}) is reduced to
\begin{eqnarray}\label{eq:mean}
{\bf P}^m(k_{\parallel},\omega)-d\chi(\omega)\bar{\bf F}_0^R(k_{\parallel},\omega){\bf P}^m(k_{\parallel},\omega)
=\chi(\omega){\bf E}^m(k_{\parallel},\omega),\nonumber\\
\end{eqnarray}
where the superscript $m$ denotes the averaged quantities
\begin{equation}
{\bf P}^m(k_{\parallel},\omega)=\frac{1}{d}\int_{-d/2}^{d/2}{\bf P}(k_{\parallel},\omega;z)dz
\end{equation}
and
\begin{equation}
{\bf E}^m(k_{\parallel},\omega)=\frac{1}{d}\int_{-d/2}^{d/2}{\bf E}(k_{\parallel},\omega;z)dz.
\end{equation}
The analysis of the explicit form of $F_{0,ij}^{R}$ (see the Appendix) reveals that $F_{0,xy}^{R}=F_{0,yx}^{R}=F_{0,yz}^{R}=F_{0,zy}^{R}=0$, while the tensorial components $F_{0,xz}^{R}$, $F_{0,zx}^{R}$ and $F_{0,zz}^{R}$ are proportional to $k_{\parallel}/\kappa$ and $(k_{\parallel}/\kappa)^2$, respectively, where 
\begin{equation}
\kappa=\sqrt{\left(\frac{\omega}{c}\right)^2\epsilon_h-k_{\parallel}^2}.
\end{equation}
and we assume that the vector ${\bf k}_{\parallel}$ is directed along the $x$ axis. Therefore in the case where $k_{\parallel}\ll \kappa$, only the components $F_{0,xx}^{R}$ and $F_{0,yy}^{R}$ contribute to Eq. (\ref{eq:mean}) in the lowest order in $d/\lambda$. In the opposite limiting case, $k_{\parallel}\gg \kappa$, the dominant contributions originate from $F_{0,zz}^{R}$ and $F_{0,yy}^{R}$. For the sake of simplicity, we assume that the diagonal elements satisfactory describe the local field in the intermediate case as well.\\
In this mean-field approximation, the linear susceptibility tensor of the slab is given by
\begin{equation}\label{eq:eta_m}
\eta^m_{jj} ( k_{\parallel},\omega )=\frac{\chi(\omega)}{1-d\chi(\omega)F_{0,jj}^R(k_{\parallel},\omega)}.
\end{equation}
Correspondingly, the mean-field dielectric tensor, which determines the nonlocal and anisotropic optical response of the slab, takes the form
\begin{equation}\label{eq:epsilon_m}
\epsilon^m_{jj}( k_{\parallel},\omega)=\epsilon_h+4\pi\eta^m_{jj}(k_{\parallel},\omega).
\end{equation}
\section{Reflection and transmission of light by a slab}\label{sec:refl_trans}
Let us assume now that a plane monochromatic electromagnetic wave of the frequency $\omega$ falls at the nanocomposite slab from medium 1 so that its wave vector ${\bf k}_1$ makes an angle $\theta$ with the normal to the boundary (Fig. \ref{fig:geometry}). We assume also that the boundary surface has a small random subwavelength roughness which leads to the scattering of the incident light. Two sources of scattering can contribute here: (i) surface roughness due to the surface morphology and (ii) "volume" roughness due to inhomogeneities in the bulk originating from the distribution of nanoparticles \cite{Raether88}. \\
The scattering cross-section is proportional to the power spectrum of the rough surface, $\mathcal{P}({\bf k}_{\parallel}-{\bf k}_{1\parallel})$, which is determined as the Fourier transform of the roughness correlation function, $C({\bf r}_{\parallel}^{\prime}-{\bf r}_{\parallel})$ \cite{Raether88,Bordo05}. For an isotropic surface one usually assumes
\begin{equation}
C(\mid{\bf r}_{\parallel}^{\prime}-{\bf r}_{\parallel}\mid)=\delta^2\exp\left(-\frac{\mid{\bf r}_{\parallel}^{\prime}-{\bf r}_{\parallel}\mid^2}{\sigma^2}\right),
\end{equation}
where $\delta$ is the root-mean-square height of the surface and $\sigma$ is the correlation length. Accordingly, 
\begin{eqnarray}\label{eq:P}
\mathcal{P}(\mid {\bf k}_{\parallel}-{\bf k}_{1\parallel}\mid)\sim G(\mid{\bf k}_{\parallel}-{\bf k}_{1\parallel}\mid)\nonumber\\
\equiv\frac{\sigma^2}{4\pi}\exp\left(-\frac{\sigma^2}{4}\mid {\bf k}_{\parallel}-{\bf k}_{1\parallel}\mid^2\right).
\end{eqnarray}
We model therefore the spatial Fourier spectrum of the incident field amplitude by the Gaussian function
\begin{equation}\label{eq:Gauss}
A_1^i({\bf k}_{\parallel},\omega)=A_0\exp\left(-\frac{\sigma^2}{8}\mid {\bf k}_{\parallel}-{\bf k}_{1\parallel}\mid^2\right),
\end{equation}
so that the corresponding intensity is proportional to the power spectrum, Eq. (\ref{eq:P}). \\
We consider separately two cases: (i) when the electric field vector of the incident wave lies in the plane of incidence (TM polarization) and (ii) when it is perpendicular to the plane of incidence (TE polarization). It should be noted that generally the scattering at a rough surface is accompanied by light depolarization. However, for scattering at small roughness in the plane of incidence the polarization of light is not changed \cite{Bordo05}. In what follows we discuss the latter case which implies that the vector ${\bf k}_{\parallel}$ for both the reflected and transmitted waves lies in the plane of incidence.\\
\subsection{TM polarization}
In this case the magnetic field vector of the incident wave is directed along the $y$ axis and its spatial Fourier components are given by
\begin{equation}
H^i_{1y}(k_{\parallel},\omega)=A_1^i(k_{\parallel},\omega)e^{ik_{\parallel}x}e^{ik_{1z}z}e^{-i\omega t}
\end{equation}
with
\begin{equation}
k_{1z}=\sqrt{\left(\frac{\omega}{c}\right)^2\epsilon_1-k_{\parallel}^2}.
\end{equation}
Similarly, the magnetic field components for the reflected wave as well as for the field inside the slab and the transmitted wave have respectively the following forms:
\begin{equation}\label{eq:Hr}
H^r_{1y}(k_{\parallel},\omega)=A_1^r(k_{\parallel},\omega)e^{ik_{\parallel}x}e^{-ik_{1z}z}e^{-i\omega t},
\end{equation}
\begin{eqnarray}\label{eq:H2}
H_{2y}(k_{\parallel},\omega)=[A_2^e(k_{\parallel},\omega)\cos k_{2z}z\nonumber\\
+A_2^o(k_{\parallel},\omega)\sin k_{2z}z]e^{ik_{\parallel}x}e^{-i\omega t}
\end{eqnarray}
and
\begin{equation}\label{eq:H3}
H_{3y}(k_{\parallel},\omega)=A_3(k_{\parallel},\omega)e^{ik_{\parallel}x}e^{ik_{3z}z}e^{-i\omega t}
\end{equation}
with
\begin{equation}\label{eq:k2z}
k_{2z}=\sqrt{\left(\frac{\omega}{c}\right)^2\epsilon^m_{xx}-k_{\parallel}^2\frac{\epsilon^m_{xx}}{\epsilon^m_{zz}}}
\end{equation}
and
\begin{equation}
k_{3z}=\sqrt{\left(\frac{\omega}{c}\right)^2\epsilon_3-k_{\parallel}^2}.
\end{equation}
Here Eq. (\ref{eq:k2z}) follows from the dispersion relation for a TM polarized field \cite{Bordo18}.\\
The corresponding electric field components are obtained from the Maxwell's equation
\begin{equation}
\nabla\times {\bf H}(k_{\parallel},\omega)=-\frac{i\omega}{c}\bar{\epsilon}(k_{\parallel},\omega){\bf E}(k_{\parallel},\omega),
\end{equation}
where the dielectric tensor $\bar{\epsilon}(k_{\parallel},\omega)$ is reduced to $\epsilon_1$ and $\epsilon_3$ for media 1 and 3, respectively, and $\epsilon_{ij}=\epsilon_{jj}^m\delta_{ij}$ for the slab. \\ 
The unknown amplitudes $A_1^r$, $A_2^e$, $A_2^o$ and $A_3$ are found from the boundary conditions at $z=\pm d/2$ which express the continuity of the tangential magnetic and electric field components. As a result, the amplitudes of the reflected and transmitted waves can be written as
\begin{equation}\label{eq:TMr}
A_1^r(k_{\parallel},\omega)=R^{TM}(k_{\parallel},\omega)A_1^i(k_{\parallel},\omega)e^{-ik_{1z}d}
\end{equation}
and 
\begin{equation}\label{eq:TMt}
A_3(k_{\parallel},\omega)=T^{TM}(k_{\parallel},\omega)A_1^i(k_{\parallel},\omega)e^{-i(k_{1z}+k_{3z})d/2},
\end{equation}
respectively. We have introduced here the following notations:
\begin{eqnarray}
R^{TM}=-\frac{D^{TM}_{e1}D^{TM}_{o3}+D^{TM}_{o1}D^{TM}_{e3}}{\mathcal{D}^{TM}},\label{eq:RTM}\\
T^{TM}=\frac{2i(k_{1z}/\epsilon_1)(D^{TM}_{e1}+D^{TM}_{o1})}{\mathcal{D}^{TM}},\label{eq:TTM}
\end{eqnarray}
where
\begin{eqnarray}
D^{TM}_{e1}=\frac{k_{2z}}{\epsilon_{xx}^m}\tan\frac{k_{2z}d}{2}-i\frac{k_{1z}}{\epsilon_1},\\
D^{TM}_{e3}=\frac{k_{2z}}{\epsilon_{xx}^m}\tan\frac{k_{2z}d}{2}+i\frac{k_{3z}}{\epsilon_3},\\
D^{TM}_{o1}=\frac{k_{2z}}{\epsilon_{xx}^m}\cot\frac{k_{2z}d}{2}+i\frac{k_{1z}}{\epsilon_1},\\
D^{TM}_{o3}=\frac{k_{2z}}{\epsilon_{xx}^m}\cot\frac{k_{2z}d}{2}-i\frac{k_{3z}}{\epsilon_3}
\end{eqnarray}
and
\begin{equation}
\mathcal{D}^{TM}=D^{TM}_{e1}D^{TM}_{o3}+D^{TM}_{o1}D^{TM}_{e3}+2i\frac{k_{1z}}{\epsilon_1}(D^{TM}_{o3}-D^{TM}_{e3}).
\end{equation}
\subsection{TE polarization}
In this case the electric field in the incident wave is directed along the $y$ axis and its Fourier components are given by
\begin{equation}
E^i_{1y}(k_{\parallel},\omega)=A_1^i(k_{\parallel},\omega)e^{ik_{\parallel}x}e^{ik_{1z}z}e^{-i\omega t}.
\end{equation}
The electric field components in the other media have the form similar to Eqs. (\ref{eq:Hr})-(\ref{eq:H3}). The corresponding magnetic field components are found from the Maxwell's equation
\begin{equation}
\nabla\times {\bf E}(k_{\parallel},\omega)=\frac{i\omega}{c}{\bf H}(k_{\parallel},\omega).
\end{equation}
Using the same boundary conditions as before one finds for the reflected and transmitted field amplitudes 
\begin{equation}\label{eq:TEr}
A_1^r(k_{\parallel},\omega)=R^{TE}(k_{\parallel},\omega)A_1^i(k_{\parallel},\omega)e^{-ik_{1z}d}
\end{equation}
and 
\begin{equation}\label{eq:TEt}
A_3(k_{\parallel},\omega)=T^{TE}(k_{\parallel},\omega)A_1^i(k_{\parallel},\omega)e^{-i(k_{1z}+k_{3z})d/2},
\end{equation}
respectively. Here the following notations have been introduced:
\begin{eqnarray}
R^{TE}=-\frac{D^{TE}_{e1}D^{TE}_{o3}+D^{TE}_{o1}D^{TE}_{e3}}{\mathcal{D}^{TE}},\label{eq:RTE}\\
T^{TE}=\frac{2ik_{1z}(D^{TE}_{e1}+D^{TE}_{o1})}{\mathcal{D}^{TE}},\label{eq:TTE}
\end{eqnarray}
where
\begin{eqnarray}
D^{TE}_{e1}=k_{2z}\tan\frac{k_{2z}d}{2}-ik_{1z},\\
D^{TE}_{e3}=k_{2z}\tan\frac{k_{2z}d}{2}+ik_{3z},\\
D^{TE}_{o1}=k_{2z}\cot\frac{k_{2z}d}{2}+ik_{1z},\\
D^{TE}_{o3}=k_{2z}\cot\frac{k_{2z}d}{2}-ik_{3z}
\end{eqnarray}
and
\begin{equation}
\mathcal{D}^{TE}=D^{TE}_{e1}D^{TE}_{o3}+D^{TE}_{o1}D^{TE}_{e3}+2ik_{1z}(D^{TE}_{o3}-D^{TE}_{e3})
\end{equation}
with
\begin{equation}
k_{2z}=\sqrt{\left(\frac{\omega}{c}\right)^2\epsilon^m_{yy}-k_{\parallel}^2}.
\end{equation}
\subsection{Reflection and transmission coefficients}
The reflected and transmitted field amplitudes for TM and TE polarizations are found as the Fourier transforms of Eqs. (\ref{eq:TMr}), (\ref{eq:TMt}) and (\ref{eq:TEr}), (\ref{eq:TEt}), respectively. The corresponding intensities are determined by the time-averaged $z$-component of the Poynting vectors
\begin{equation}\label{eq:S}
<S^{\mu}_{z}({\bf r},\omega)>=\frac{c}{8\pi}\text{Re}[E_{jx}^{\mu}({\bf r},\omega)H_{jy}^{\mu *}({\bf r},\omega)], \quad \mu=i,r,t,
\end{equation}
where $j=1$ for the incident $(i)$ and reflected ($r$) fields and $j=3$ for the transmitted ($t$) field.\\
Assuming that the incident light beam diameter is much larger than the surface roughness correlation length, the observable reflectivity and transmissivity are found through the integrals over the whole slab surface as
\begin{eqnarray}\label{eq:refl}
\mathcal{R}^{\alpha}(\omega)=\frac{\int <S^r_z({\bf r},\omega)> d{\bf r}_{\parallel}}{\int <S^i_z({\bf r},\omega)> d{\bf r}_{\parallel}}\nonumber\\
=\int_{-\infty}^{\infty} \text{Re}(k_{1z})\mid R^{\alpha}(k_{\parallel},\omega)\mid^2G(\mid k_{\parallel}-k_{1\parallel}\mid)dk_{\parallel}\nonumber\\
\times \left[\int_{-\infty}^{\infty} \text{Re}(k_{1z})G(\mid k_{\parallel}-k_{1\parallel}\mid)dk_{\parallel}\right]^{-1}
\end{eqnarray}
and
\begin{eqnarray}\label{eq:trans}
\mathcal{T}^{\alpha}(\omega)=\frac{\int <S^t_z({\bf r},\omega)> d{\bf r}_{\parallel}}{\int <S^i_z({\bf r},\omega)> d{\bf r}_{\parallel}}\nonumber\\
=\delta^{\alpha}\int_{-\infty}^{\infty} \text{Re}(k_{3z})\mid T^{\alpha}(k_{\parallel},\omega)\mid^2G(\mid k_{\parallel}-k_{1\parallel}\mid)dk_{\parallel}\nonumber\\
\times \left[\int_{-\infty}^{\infty} \text{Re}(k_{1z})G(\mid k_{\parallel}-k_{1\parallel}\mid)dk_{\parallel}\right]^{-1},\nonumber\\
\end{eqnarray}
respectively, where the superscript $\alpha$ denotes either TM or TE polarization, $\delta^{TM}=\epsilon_1/\epsilon_3$, $\delta^{TE}=1$ and $k_{1\parallel}=(\omega/c)\sqrt{\epsilon_1}\sin\theta$. Let us note that the ranges of integration in Eqs. (\ref{eq:refl}) and (\ref{eq:trans}) are actually limited by the conditions $\mid k_{\parallel}\mid \le (\omega/c)\sqrt{\epsilon_1}$ and $\mid k_{\parallel}\mid \le(\omega/c)\sqrt{\epsilon_3}$.\\
\section{Numerical results}\label{sec:numer}
We illustrate the above theory with some numerical calculations carried out for a model system which simulates the structure experimentally investigated in Ref. \cite{Elbahri11}. Namely, we assume that Au NPs of radius $R=2$ nm are randomly distributed in the dielectric host medium with $\epsilon_h=1.5^2$, the light beam is incident from vacuum ($\epsilon_1=1$) and the nanocomposite slab is supported by an optically thick gold film so that the transmission through the structure is zero. In such a case, assuming that the light scattered in the slab is completely absorbed as well, the absorptivity, $\mathcal{A}^{\alpha}$, can be expressed in terms of the reflectivity as $\mathcal{A}^{\alpha}=1-\mathcal{R}^{\alpha}$. The dielectric function of the gold film, $\epsilon_3(\omega)$, is taken in the Drude model, Eq. (\ref{eq:drude}), with the parameters $\epsilon_{\infty}=9$, $\omega_p=13.8\times 10^{15}$ s$^{-1}$ and the relaxation constant $\Gamma=0.11\times 10^{15}$ s$^{-1}$ \cite{Shalaev10}. For gold NPs $\Gamma$ should be replaced by $\Gamma_{NP}$ which results not only from the electron collision rate given by $\Gamma$, but also from the electron collisions with the NP boundary, so that $\Gamma_{NP}\approx\Gamma+v_F/R$ with $v_F$ being the Fermi velocity \cite{Shalaev10}. For Au NPs of radius $R=2$ nm this gives $\Gamma_{NP}\approx 0.81\times 10^{15}$ s$^{-1}$. \\
For dense NP arrays which were realized in Ref. \cite{Elbahri11} we envisage, however, additional channels of LSP relaxation.  A hint for that can be found, for example, in the optical absorbance spectrum of diamond-like nanocomposite films containing Ag NPs which demonstrates broadening with the increase of the NP concentration \cite{Tamulevicius14}. One can propose two mechanisms responsible for such behavior as follows.\\
In the quasistatic approximation, which is valid for NPs of subwavelength diameters, the LSP frequencies of different spherical NPs coincide with each other. For small enough distances between NPs this can lead to "jumps" of LSP excitations from one NP to another which disrupt the collective electron oscillations in a single NP and reduce the LSP relaxation time \cite{Bordo19}.\\
Another possible reason for the enhancement of the LSP decay rate is the cooperative radiation of NPs. In the model under discussion the nanocomposite slab cab be regarded as an ensemble of point dipole emitters which is confined within a subwavelength layer. Such a system is known to radiate coherently with a decay rate enhanced by the factor $\eta=Nd(\lambda/2\pi)^2$ (collective spontaneous emission or Dicke superradiance) \cite{Glauber00}. The radiative damping rate of a single spherical NP can be estimated using the formula \cite{Bordo19,Melikyan04}
\begin{equation}\label{eq:rad}
\Gamma_{rad}=\frac{2}{9}\frac{\sqrt{\epsilon_h}}{2\epsilon_h+1}\frac{\omega_p^4R^3}{c^3}.
\end{equation}
Taking here $R=2$ nm, $f=0.1$, $d=50$ nm and $\lambda=500$ nm one obtains for gold NPs $\eta\Gamma_{rad}\approx 0.61\times 10^{15}$ s$^{-1}$ that is comparable with the estimate for $\Gamma_{NP}$ given above. The contributions from different relaxation channels are added together resulting in the total relaxation rate. Therefore in what follows we consider the quantity $\Gamma_{NP}$ as an empirical parameter and investigate its influence on the optical spectra.
\subsection{Maxwell Garnett approximation}
\begin{figure}
\includegraphics[width=\linewidth]{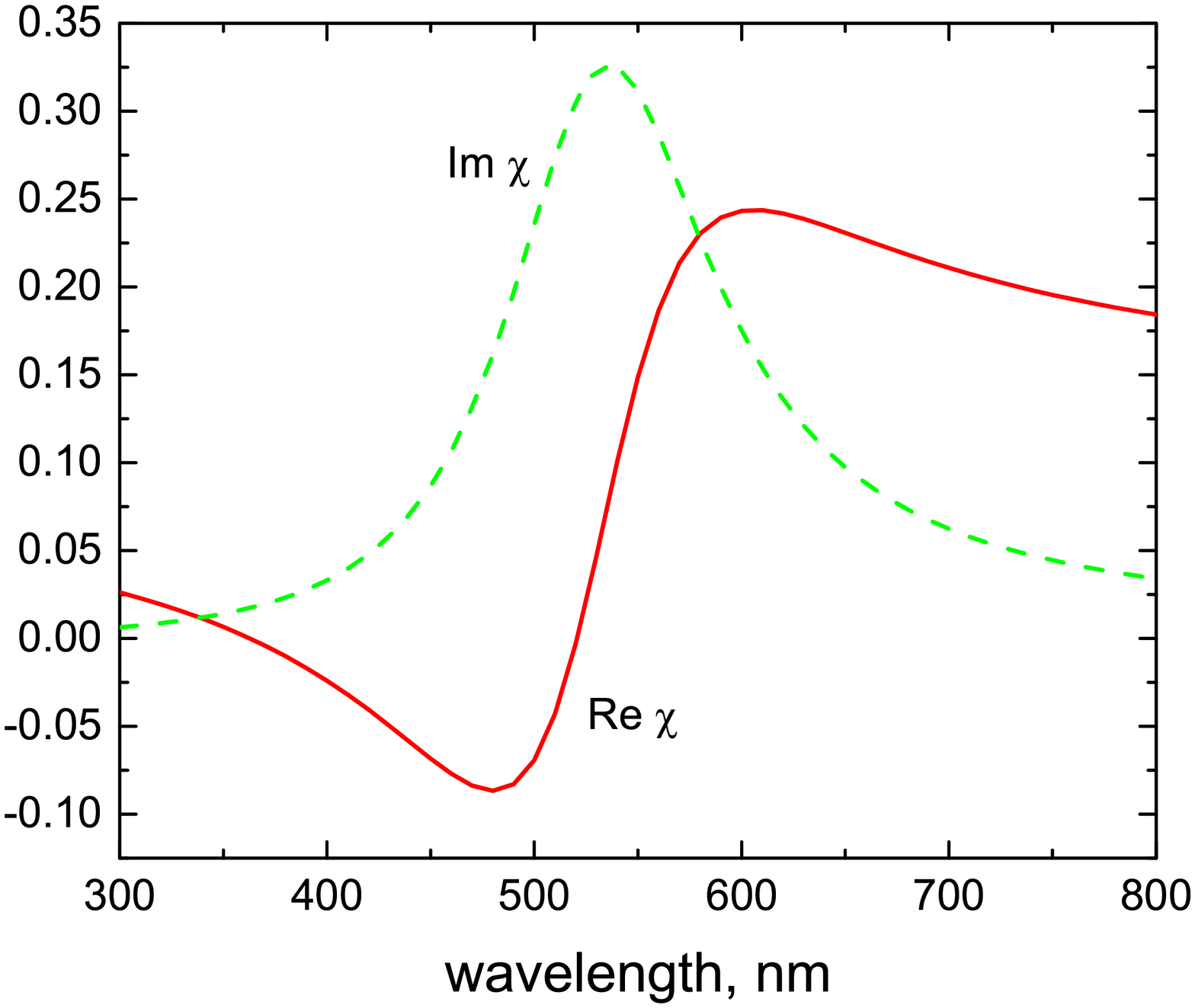}
\caption{\label{fig:chi} The spectral dependence of the real and imaginary parts of the linear susceptibility of Au NPs. The Maxwell Garnett approximation, $\Gamma_{NP}=0.81\times 10^{15}$ s$^{-1}$ and $f=0.2$.}
\end{figure}

To elucidate the role of different factors on the reflection spectrum, we calculate it first in the Maxwell Garnett (MG) approximation which corresponds to the neglect of reflections from the nanocomposite slab boundaries ($F^R_{0,jj}\equiv 0$). We neglect also at this point the effect of the LSP resonance broadening due to the interaction between NPs ($\Gamma_{NP}= 0.81\times 10^{15}$ s$^{-1}$) and the light scattering at surface roughness [$G(\mid k_{\parallel}-k_{1\parallel}\mid)\rightarrow \delta(k_{\parallel}-k_{1\parallel})$]. The linear susceptibility of Au NPs embedded in the host medium, Eq. (\ref{eq:chi}), is plotted as a function of the wavelength in Fig. \ref{fig:chi}. Its spectral dependence indicates the frequency range between 450 and 650 nm as the LSP absorption band.\\
\begin{figure}
\includegraphics[width=\linewidth]{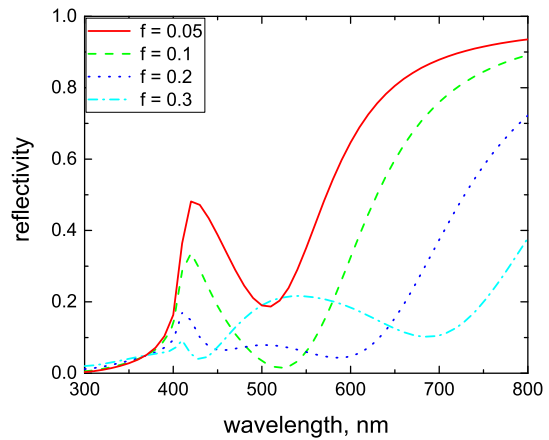}
\caption{\label{fig:TM_fraction} The dependence of the reflection spectrum in TM polarization on the volume fraction of Au NPs in the slab. The Maxwell Garnett approximation, $d=50$ nm, $\Gamma_{NP}=0.81\times 10^{15}$ s$^{-1}$, $\theta=20^{\circ}$.}
\end{figure} 
\begin{figure}
\includegraphics[width=\linewidth]{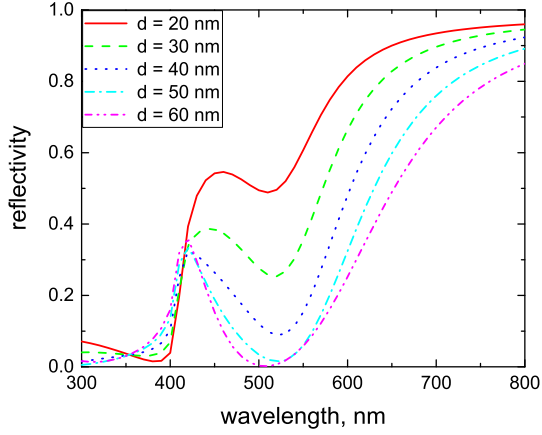}
\caption{\label{fig:TM_MG_thickness} The dependence of the reflection spectrum in TM polarization on the slab thickness. The Maxwell Garnett approximation, $\Gamma_{NP}=0.81\times 10^{15}$ s$^{-1}$, $f=0.1$ and $\theta=20^{\circ}$.}
\end{figure}
\begin{figure}
\includegraphics[width=\linewidth]{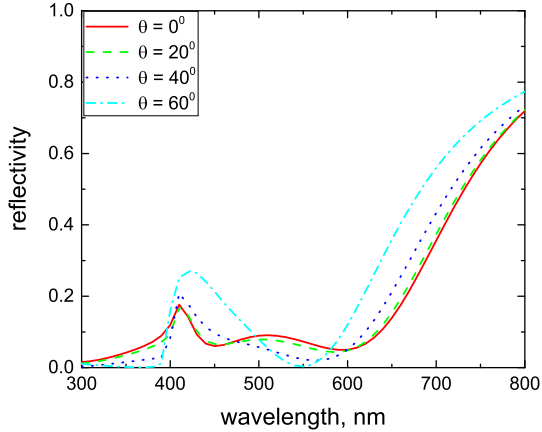}
\caption{\label{fig:TM_MG_angles} The dependence of the reflection spectrum in TM polarization on the angle of incidence. The Maxwell Garnett approximation, $d=50$ nm, $\Gamma_{NP}=0.81\times 10^{15}$ s$^{-1}$ and $f=0.2$.}
\end{figure}
Figure \ref{fig:TM_fraction} shows the reflectivity in TM polarization as a function of the volume fraction of NPs, $f=(4\pi/3)R^3N$, calculated in the MG approximation. One can see that the average value of $\mathcal{R}^{TM}$ in the spectral range which corresponds to the LSP band depends non-monotonically on $f$. It has a minimum at $f=0.2$ and increases for both $f<0.2$ and $f>0.2$. Let us note that the existence of an optimum value of the NP filling factor for perfect absorption was also reported in Ref. \cite{Elbahri11}. The dependence of the reflectivity on the slab thickness is represented in Fig. \ref{fig:TM_MG_thickness}, while its dependence on the angle of incidence is given in Fig. \ref{fig:TM_MG_angles}.\\
\begin{figure}
\includegraphics[width=\linewidth]{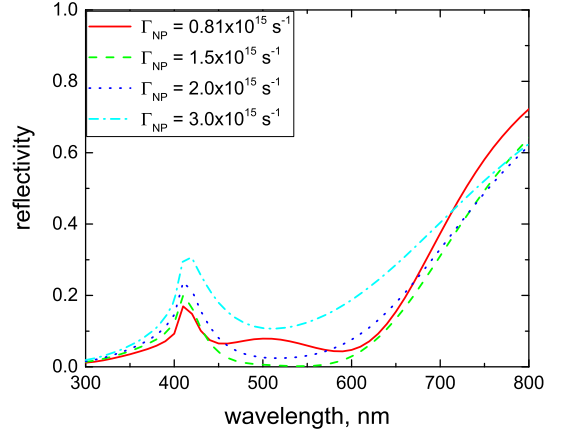}
\caption{\label{fig:TM_interaction} The dependence of the reflection spectrum in TM polarization on $\Gamma_{NP}$. $d=50$ nm, $f=0.2$ and $\theta=20^{\circ}$.}
\end{figure} 
\begin{figure}
\includegraphics[width=\linewidth]{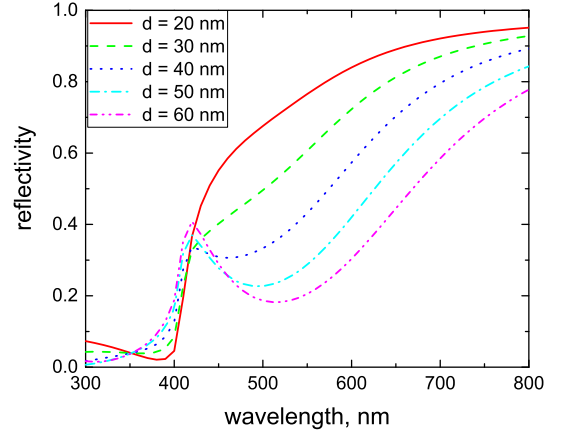}
\caption{\label{fig:TM_thickness} The dependence of the reflection spectrum in TM polarization on the slab thickness. $\Gamma_{NP}=2.0\times 10^{15}$ s$^{-1}$, $f=0.1$ and $\theta=20^{\circ}$.}
\end{figure}
\subsection{Beyond the Maxwell Garnett approximation}
\begin{figure}
\includegraphics[width=\linewidth]{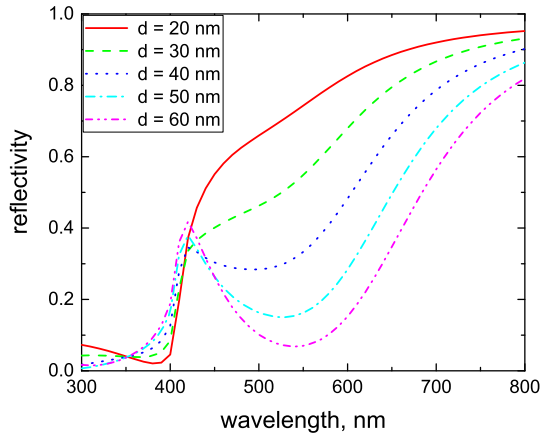}
\caption{\label{fig:TM_reflections_thickness} The dependence of the reflection spectrum in TM polarization on the slab thickness with account of reflections from the slab boundaries. $\Gamma_{NP}=2.0\times 10^{15}$ s$^{-1}$, $f=0.1$ and $\theta=20^{\circ}$.}
\end{figure}

Figure \ref{fig:TM_interaction} illustrates how the reflectivity $\mathcal{R}^{TM}$ calculated in the MG approximation for the optimum value depends on the LSP relaxation rate, $\Gamma_{NP}$. Again one observes a non-monotonic behavior with the broadband minimum in reflectivity at $\Gamma_{NP}=1.5\times 10^{15}$ s$^{-1}$. Figure \ref{fig:TM_thickness} demonstrates how the increased value of $\Gamma_{NP}$ dramatically modifies the thickness dependence of the reflectivity leading to the extension of the efficient absorption frequency band with increase in $d$ (compare with Fig. \ref{fig:TM_MG_thickness}). Figure \ref{fig:TM_reflections_thickness} shows the influence of reflections from the slab boundaries. One can see that they do not change the overall spectral dependence of the reflectivity, but reduce however it noticeably within the LSP frequency band.\\
\begin{figure}
\includegraphics[width=\linewidth]{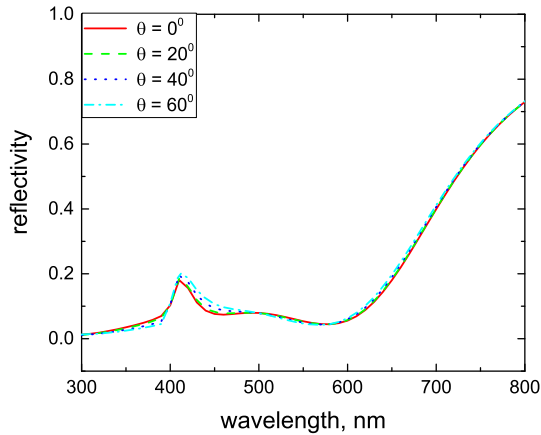}
\caption{\label{fig:TM_scattering} The dependence of the reflection spectrum in TM polarization on the angle of incidence with account of light scattering at the slab surface. $d=50$ nm, $\Gamma_{NP}=0.81\times 10^{15}$ s$^{-1}$, $f=0.2$ and $\sigma=200$ nm.}
\end{figure}
\begin{figure}
\includegraphics[width=\linewidth]{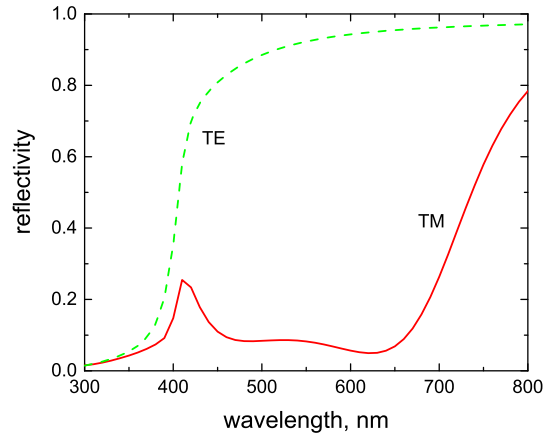}
\caption{\label{fig:TM_TE} The reflection spectrum for TM and TE polarizations with account of all factors together. $d=50$ nm, $\Gamma_{NP}=2.0\times 10^{15}$ s$^{-1}$, $f=0.2$, $\theta=20^{\circ}$ and $\sigma=100$ nm.}
\end{figure}
\begin{figure}
\includegraphics[width=\linewidth]{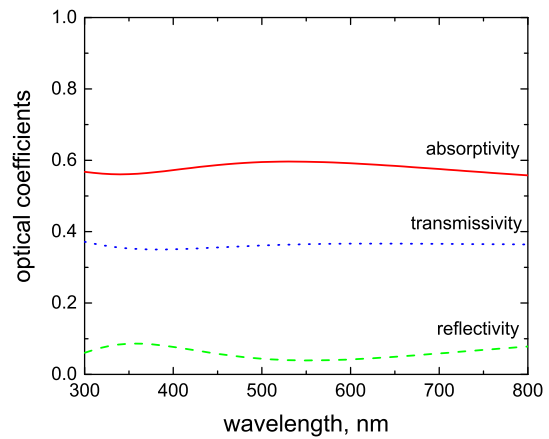}
\caption{\label{fig:Al2O3} The reflection, transmission and absorption spectra for TM polarization for a nanocomposite Ag-Al$_2$O$_3$ between vacuum and a glass substrate. $d=150$ nm, $\Gamma_{NP}= 17\times 10^{15}$ s$^{-1}$, $f=0.2$, $\theta=0^{\circ}$ and $\sigma=100$ nm.}
\end{figure}
To examine the influence of light scattering at the vacuum-slab interface, we have calculated the reflectivity spectrum, Eq. (\ref{eq:refl}), for the surface roughness correlation length typical for surfaces of small roughness \cite{Raether88}. Figure \ref{fig:TM_scattering} demonstrates an omnidirectional character of perfect absorption in this case for TM polarization in contrast with the angle dependence in the MG approximation shown in Fig. \ref{fig:TM_MG_angles}. A similar angle insensitivity is observed for TE polarization (not shown). Such behavior can be understood as follows. Due to light scattering the spatial Fourier spectrum of the light penetrating into the slab, Eq. (\ref{eq:Gauss}), is very broad which means that it contains wave vectors corresponding to a large range of incidence angles. As a result, the reflection spectrum is insensitive to the direction of the wave vector of light before scattering.\\
Finally, we have calculated the reflection spectra for both polarizations taking into account all the factors examined above together. The results for optimum values of parameters which provide a minimum reflection are shown in Fig. \ref{fig:TM_TE}. The spectrum for TM polarization demonstrates a broad band of low reflectivity (high absorptivity). On the contrary, the spectrum for TE polarization displays no broadband absorption. This result points at a strong depolarization in the light scattering at surface roughness as a possible reason of the polarization insensitivity observed in experiments \cite{Grigorenko10,Elbahri11}.\\
The plot shown in Fig. \ref{fig:TM_TE} for TM polarization can be compared with the experimental data reported in Ref. \cite{Elbahri11}. There the reflectivity of the structure consisting of a glass substrate coated with an optically thick Au film followed by a 25 nm SiO$_2$ layer and a 20 nm film of nanocomposite Au/SiO$_2$ was investigated. The optimum nanocomposite filling factor, which provided a maximum absorption, was 40\%. As far as the host material in the nanocomposite is the same as that of the spacer layer, there are no reflections of the dipole field at the boundary between them. Therefore to apply the approach developed in this paper one can recalculate the filling factor to the total thickness of the two layers (45 nm) that gives the volume fraction $f\approx 0.18$ very close to the optimum value in Fig. \ref{fig:TM_TE}. The reflectivity reported in Ref. \cite{Elbahri11} for the spectral range 400 - 750 nm does not exceed 10\% and reaches its minimum of 0 - 5$\%$, depending on the angle of incidence, in the range 500 - 650 nm. This tendency is satisfactory reproduced by Fig. \ref{fig:TM_TE} besides a peak at 420 nm.\\
We have also applied the approach developed in this paper to the nanocomposite structure investigated in Ref. \cite{Grigorenko10}. There a broadband ($240-850$ nm) absorption at the level of above 90\% was experimentally demonstrated for a nanocomposite film of thickness $d=150-160$ nm which contained Ag NPs of radii $R=50-60$ nm randomly distributed with volume fractions $0.08<f<0.3$. The host material was Al$_2$O$_3$ ($\epsilon_h=1.7^2$) and the film was deposited onto a glass substrate ($\epsilon_3=1.5^2$).\\ 
For such large NPs the dominant LSP relaxation channel is radiative damping. Using formula (\ref{eq:rad}) one obtains for Ag NPs ($\omega_p=14.0\times 10^{15}$ s$^{-1}$ \cite{Shalaev10}) of radius $R=60$ nm $\Gamma_{NP}\approx\Gamma_{rad}=17\times 10^{15}$ s$^{-1}$. The reflection, transmission and absorption spectra for this case are shown in Fig. \ref{fig:Al2O3}. Although one cannot expect a quantitative agreement with our model which assumes that the NP sizes are much smaller than the slab thickness, nevertheless it reproduces the broadband character of absorption.
\subsection{Bruggeman approximation}
\begin{figure}
\includegraphics[width=\linewidth]{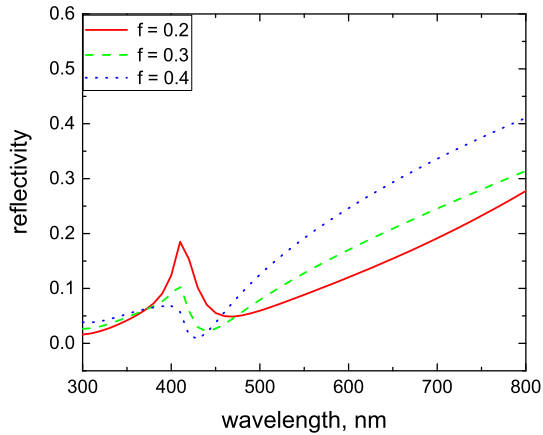}
\caption{\label{fig:BG} The reflection spectrum calculated in the Bruggeman model for different NP volume fractions. $d=50$ nm, $\Gamma_{NP}=0.81\times 10^{15}$ s$^{-1}$ and $\theta=0^{\circ}$.}
\end{figure}

It is interesting to compare the results obtained above with the Bruggeman model which can be applied, at least formally, for nanocomposites with arbitrary volume fractions of inclusions. In this model the effective dielectric function of the slab is given by \cite{Markel16}
\begin{equation}
\epsilon_2(\omega)=\frac{1}{4}\left[b(\omega)+\sqrt{8\epsilon_h\epsilon(\omega)+b(\omega)^2}\right],
\end{equation}
where
\begin{equation}
b(\omega)=(2f_h-f)\epsilon_h+(2f-f_h)\epsilon(\omega)
\end{equation}
and $f_h=1-f$.\\
The reflection spectra calculated in the Bruggeman model using the Fresnel equations for normal incidence and different volume fractions are shown in Fig. \ref{fig:BG}. One can see that a broadband reduced reflection does not emerge even at relatively large volume fractions of NPs.
This result reveals that a large filling factor itself does not lead to broadband perfect absorption and probably the mechanisms of the decay rate enhancement discussed above play a crucial role in this phenomenon.
\section{Conclusion}\label{sec:concl}
In this paper, we have developed the theory of the optical properties of an optically thin nanocomposite slab which contains randomly distributed metal nanoparticles. We have derived the analytical formulas for the reflection and transmission coefficients in both polarizations, Eqs. (\ref{eq:RTM}), (\ref{eq:TTM}), (\ref{eq:RTE}) and (\ref{eq:TTE}), which take into account the anisotropy and nonlocality of the slab optical response. In the developed theory the effective linear susceptibility of the nanocomposite slab is determined by the number density of NPs, while their positions are unimportant. This means that the approach under discussion is equally applicable to plasmonic metamaterials with a regular arrangement of spherical metal NPs.\\ 
We have examined the impact of different factors, which are beyond the scope of the Maxwell Garnett approximation, on the reflection spectrum of a plasmonic nanocomposite slab placed on a metal substrate. We have concluded that the broadband perfect absorption in such structures is associated with the LSP absorption band and originates from both reflections from the slab boundaries, which trap light within the slab, and enhanced decay channels for the LSP excitations in a dense NP array. In contrast to this, the omnidirectional character of perfect absorption and its polarization insensitivity stem from the light scattering and depolarization, respectively, at surface roughness or volume irregularities in the slab. \\
The reflection spectra calculated with the use of the developed approach satisfactory agree with the available experimental data, while those obtained in both Maxwell Garnett and Bruggeman approximations do not. This conclusion reveals the importance of a proper account of the local field in the slab and adequate modeling of the LSP relaxation in dense NP arrays.\\ 
These findings open up new possibilities for engineering broadband perfect absorption in plasmonic nanocomposites and can stimulate experimental investigations in this direction. The LSP relaxation rate is determined by both the NP filling factor and NP sizes, while the light scattering stems from surface morphology - the factors which can be well controlled in an experiment. The perfect absorption band can be extended to longer wavelengths when using metal nanorods with large aspect ratios as NPs in the nanocomposite \cite{Pelton08}.\\
Another important aspect which is highlighted by the present research is the mechanisms of the LSP relaxation in dense NP arrays which have not been discussed before. Both quantum jumps of LSP excitations between NPs and their cooperative radiation are processes which are inherent to quantum emitters. The experimental investigation of these mechanisms would therefore contribute to the rapidly developing field of quantum plasmonics \cite{Tame13}.\\
Finally, it is necessary to stress that the approach developed in this paper is a fully analytical one which has been derived from first principles. Although it does not substitute the numerical methods in all respects, it can serve as a reliable test for them. The numerical calculations should always be accompanied by a careful analysis of adopted approximations and possible numerical errors, not saying about a necessary compromise between the accuracy and the required computational memory and time. They have therefore their own range of applicability. The usual practice is testing of a computational package against known analytical solutions \cite{Draine08}. In this context, any analytical solution derived from first principles is extremely important. 
\section*{Acknowledgments}
This work was carried out under a programme of, and funded by, the European Space Agency (ESA Contract No. 4000130078/20/NL/SC). The author is grateful to Prof. Sigitas Tamulevi{\v c}ius and Prof. Tomas Tamulevi{\v c}ius for useful discussions.\\
\section*{Appendix: Explicit form of the quantities $F_{0,ij}^{R}$}
The components of the tensor $\bar{F}_0^{R}(k_{\parallel},\omega)$, Eq. (\ref{eq:F0R}), have the following form:
\begin{eqnarray} 
F^R_{0,xx}=-\frac{2\pi i\kappa}{\epsilon_h}\frac{r_1^p(1-r_2^p)+r_2^p(1-r_1^p)}{1-r_1^pr_2^p},\\
F^R_{0,yy}=\frac{2\pi i\omega^2}{\kappa c^2}\frac{r_1^s+r_2^s+2r_1^sr_2^s}{1-r_1^sr_2^s},\\
F^R_{0,zz}=\frac{2\pi ik_{\parallel}^2}{\epsilon_h\kappa}\frac{r_1^p(1+r_2^p)+r_2^p(1+r_1^p)}{1-r_1^pr_2^p},\\
F^R_{0,xz}=-F^R_{0,zx}=\frac{2\pi ik_{\parallel}}{\epsilon_h}\frac{r_2^p-r_1^p}{1-r_1^pr_2^p},\\
F^R_{0,xy}=F^R_{0,yx}=F^R_{0,yz}=F^R_{0,zy}=0.
\end{eqnarray}
Here 
\begin{eqnarray}
r_1^p=\frac{\epsilon_1 \kappa-\epsilon_hk_{1z}}{\epsilon_1 \kappa+\epsilon_hk_{1z}},\\
r_1^s=\frac{\kappa-k_{1z}}{\kappa+k_{1z}},\\
r_2^p=\frac{\epsilon_3 \kappa-\epsilon_hk_{3z}}{\epsilon_3 \kappa+\epsilon_hk_{3z}},\\
r_2^s=\frac{\kappa-k_{3z}}{\kappa+k_{3z}}
\end{eqnarray}
are the Fresnel reflection coefficients for $p$ and $s$ polarizations at the medium 1/medium 2 and medium 2/medium 3 interfaces, respectively, in the approximation of small NP volume fraction.

\end{document}